\documentclass[aps, prl, twocolumn, 
amssymb,superscriptaddress]{revtex4}
\usepackage{graphicx, bm, amsmath, amsfonts,amssymb}

\usepackage{subfigure}

\def\v#1{\mathbf{#1}}

\def\t#1{\widetilde{#1}}

\def\up{\uparrow}
\def\down{\downarrow}

\begin{document}

\title{Neutrino as topological Majorana zero modes: the origin of three generations of neutrinos and their mass mixing}

\author{Zheng-Cheng Gu}
\affiliation{Perimeter Institute for Theoretical Physics, Waterloo, Ontario, N2L2Y5, Canada}

\begin{abstract}
Recently, Majorana's spirit returns to modern condensed matter physics -- in the context of topological Majorana zero mode that represents a local half degree of freedom carrying non-Abelian statistics.
In this paper, we investigate the topological nature of a Majorana fermion by assuming that it is made up of four Majorana zero modes at cutoff energy scale. First, we show that a pair of Majorana zero modes can realize a $T^4=-1$ time reversal symmetry, a $P^4=-1$ parity symmetry and even a nontrivial $\overline C^4=-1$ charge conjugation symmetry.
Next, we propose a $\overline CPT$ super algebra for the Majorana fermion made up of four Majorana zero modes. Furthermore, the origin of three generations of neutrinos(assuming they are Majorana fermions) can be naturally explained as three distinguishable ways to form a pair of (local) complex fermions out of four Majorana zero modes. Finally, we compute the neutrino mass mixing matrix and mass ratios of the three mass eigenstates from a first principle at leading order(in the absence of $CP$ violation and charged lepton corrections). We obtain $\theta_{12}=31.7^\circ, \theta_{23}=45^\circ$, $\theta_{13}=0^\circ$ and $m_1/m_3=m_2/m_3=3/\sqrt{5}$. We predict the effective mass in neutrinoless double beta decay to be $m_{\beta\beta}=m_1/\sqrt{5}$.
\end{abstract}

\maketitle

{\it Introduction} ---
The neutrino, first discovered in 1956\cite{neutrino1}, has extremely weak interactions with other particles and therefore is named as the "ghost particle".  Since neutrinos and antineutrinos are neutral particles, it has been proposed that they are actually the same particle and can be described by a four component real Lorentz spinor---the Majorana fermion\cite{Majorana}.
The smoking gun experiment that can confirm the Majorana fermion nature of a neutrino is the so-called neutrinoless double-$\beta$ decay. Unfortunately, a definite experimental evidence is still missing so far.

A cutting-edge step towards understanding the mysteries of neutrinos has been taken by the neutrino oscillation experiments during the past decade\cite{neutrinoexp1,neutrinoexp2,neutrinoexp3,neutrinoexp4,neutrinoexp5,neutrinoexp6,neutrinoexp7,neutrinoexp8,neutrinoexp9,neutrinoexp10}. These experiments have confirmed that there are three generations of neutrinos with nonzero masses at the energy scale of $0.1eV$.
Although the Standard Model(SM) predicts a zero neutrino mass, one way to explain the origin of neutrino mass is the so called seesaw mechanism in the extended SM\cite{seesaw1,seesaw2,seesaw4}---by introducing a heavy, massive sterile right-handed neutrino that does not carry any electric-weak charge, a small mass for the left-handed light neutrino can be induced. However, this theory does not explain why there are three generations of neutrinos and where do those mystery mixing angles come from.

Recently, Majorana's spirit returns to modern condensed matter physics \cite{WilczekMajorana}---in the context of topological Majorana zero mode that arises in certain classes of topological superconductors(TSCs)\cite{KitaevMajorana,ReadMajorana}. It has been shown that a topological Majorana zero mode carries non-Abelian statistics and has a quantum dimension $\sqrt{2}$\cite{KitaevMajorana,ReadMajorana}, thus it represents a local half degree of freedom. Searching for Majorana zero mode has become a fascinating subject both theoretically\cite{SankaMajorana,Majoranatheory1,Majoranatheory2} and experimentally\cite{Majoranaexp1,Majoranaexp2,Majoranaexp3}.

In this paper, we follow the seesaw mechanism idea which requires neutrino to be a Majorana fermion\cite{Majorananeutrino}, and attempt to investigate the topological nature of neutrinos by assuming that a relativistic Majorana fermion is made up of four Majorana zero modes at cutoff energy scale. We begin with an exactly solvable $1$D condensed matter model which realizes a $T^2=-1$ time reversal symmetry protected TSC and show that the pair of Majorana zero modes on its ends realize a $T^4=-1$ fractionalized time reversal symmetry. We then show that a pair of Majorana zero modes can also realize a $P^4=-1$ parity symmetry and even a nontrivial $\overline C^4=-1$ charge conjugation symmetry. These fractionalized $\overline C,P,T$ symmetries motivate us to define a $\overline CPT$ super algebra for a Majorana fermion made up of four Majorana zero modes. Interestingly, we find that the nontrivial charge conjugation symmetry $\overline C$ changes the sign of the Majorana mass term.(It is well known that the usual charge conjugation symmetry $C$ has a trivial action on a real Majorana fermion.) Therefore, we propose that $\overline C$ is indeed a $\mathbb{Z}_2$ gauge symmetry and its spontaneous breaking(through the Anderson-Higgs mechanism) leads to the origin of right handed sterile neutrino mass.
These new concepts can even explain the origin of three generations of neutrinos, as out of four Majorana zero modes, there are three inequivalent ways to form a pair of (local) complex fermions, each characterized by the $T^4=-1$, $(TP)^4=-1$ or $(T\overline C)^4=-1$ fractionalized symmetries that it carries. Together with the $\mathbb{Z}_2$ gauge (minimal coupling) principle, we can completely determine the neutrino mass mixing matrix at leading order(LO)(without CP violation and charged lepton contributions).

\begin{figure}[t]
\begin{center}
\vskip -1.0cm
\includegraphics[width=7cm]{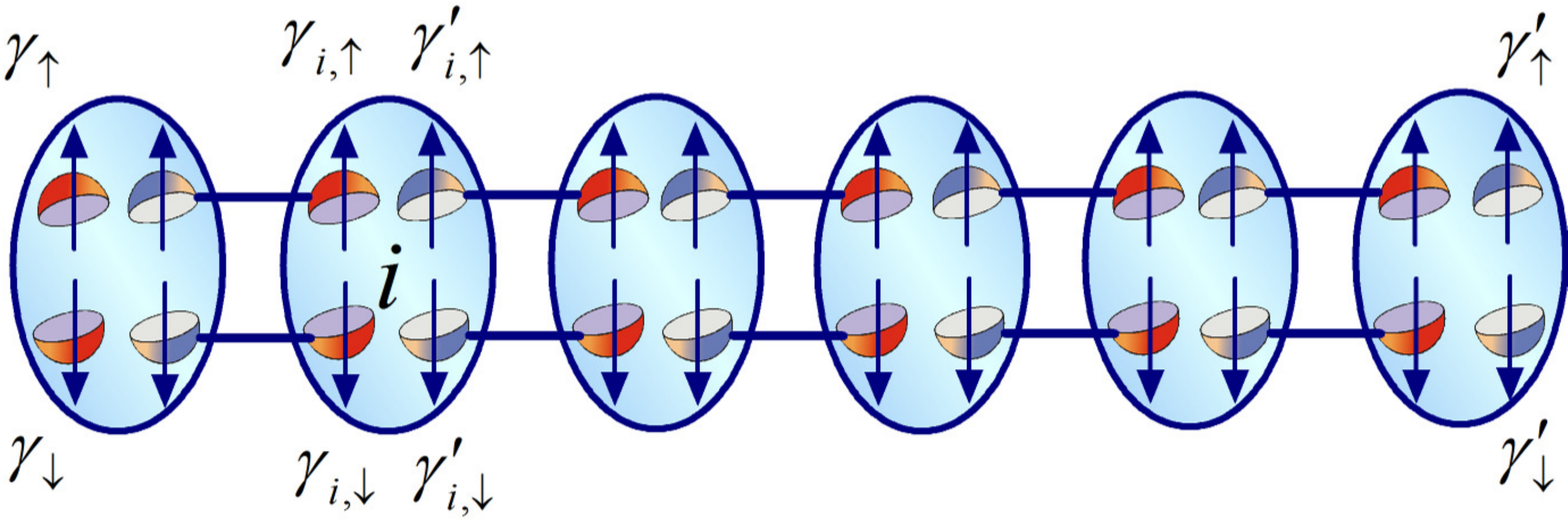}
\vskip -1.5cm
\caption{(color online)A $1$D topological superconductor protected by the $T^2=-1$ time reversal symmetry can be constructed by two copies of Kitaev's Majorana chains with opposite spin species.
The pair of dangling Majorana zero modes on left/right end are protected by the time reversal symmetry.
}\label{chain}
\end{center}
\vskip -0.5cm
\end{figure}
{\it $T^4=-1$ time reversal symmetry for a pair of Majorana zero modes} ---
To begin, we consider a $1$D topological superconductor protected by $T^2=-1$ time reversal symmetry(known as $\rm{DIII}$ class)\cite{Ryuperiod,Kitaevperiod}, which realizes a special symmetry protected topological(SPT) phase\cite{GuSPT} in $1$D.
The simplest model to realize such a $1$D TSC is just two copies of Kitaev's Majorana chains\cite{KitaevMajorana} with opposite spin species, as seen in Fig. \ref{chain}, described by the Hamiltonian
$H=\sum_{i=1}^{N-1}\sum_{\sigma} i\sigma\gamma_{i,\sigma}^{\prime}\gamma_{i+1,\sigma}$,
where Majorana operators $\gamma_{i,\sigma}$ and $\gamma_{i,\sigma}^\prime$ satisfy
$\{\gamma_{i,\sigma},\gamma_{i^\prime,\sigma^\prime}^\prime\}=0$, $\{\gamma_{i,\sigma},\gamma_{i^\prime,\sigma^\prime}\}=2\delta_{ii^\prime}\delta_{\sigma\sigma^\prime}$ and $\{\gamma_{i,\sigma}^\prime,\gamma_{i^\prime,\sigma^\prime}^\prime\}=2\delta_{ii^\prime}\delta_{\sigma\sigma^\prime}$.
Under time reversal symmetry, the Majorana operators $(\gamma_{i,\up},\gamma_{i,\down})$ and $(\gamma_{i,\up}^\prime,\gamma_{i,\down}^\prime)$ transform as:
\begin{eqnarray}
T \gamma_{i,\uparrow} T^{-1}=
-\gamma_{i,\downarrow}; \quad
T \gamma_{i,\downarrow} T^{-1}= \gamma_{i,\uparrow} \label{T1}\\
T \gamma_{i,\uparrow}^\prime T^{-1}=
-\gamma_{i,\downarrow}^\prime; \quad
T \gamma_{i,\downarrow}^\prime T^{-1}= \gamma_{i,\uparrow}^\prime,
\label{T2}
\end{eqnarray}

As seen in Fig. \ref{chain},  because the fermion mass terms
$i\gamma_{\uparrow}\gamma_{\downarrow}$ and $i\gamma_{\uparrow}^\prime\gamma_{\downarrow}^\prime$ change sign under the time reversal symmetry, a pair of dangling Majorana modes with opposite spins becomes zero modes( $\gamma_{\uparrow}\equiv\gamma_{1,\uparrow},\gamma_{\downarrow}\equiv\gamma_{1,\downarrow}$ for the left end and
$\gamma_{\uparrow}^\prime\equiv\gamma_{N,\uparrow}^\prime,\gamma_{\downarrow}^\prime\equiv\gamma_{N,\downarrow}^\prime$ for the right end)and is stable against $T$-preserving perturbations.
Recent progress in the classification of $1D$ SPT phases\cite{XieSPT1,XieSPT2} further pointed out that the edge Majorana zero modes of DIII class TSC indeed carry the $T^4=-1$ projective representation of time reversal symmetry.

A simple reason why a pair of Majorana zero modes carries a fractionalized $T^4=-1$ time reversal symmetry can be explained as the following:
The local Hilbert space on a single site for the above $T^2=-1$ TSC is a Fock-space involving both fermion parity odd sector $c_{i,\up}^\dagger |0\rangle,c_{i,\down}^\dagger|0\rangle$ and parity even sector $|0\rangle,c_{i,\up}^\dagger c_{i,\down}^\dagger|0\rangle$.( $c_{i,\sigma}=\gamma_{\sigma}+ i\gamma_{\sigma}^\prime$ are the corresponding complex fermion operators) The fermion parity odd sector satisfies $T^2=-1$ while the fermion parity even sector satisfies $T^2=1$. As a result, the time reversal symmetry group for many body fermion systems has been extended over the $\mathbb{Z}_2$ fermion parity symmetry group $\{I,P_f\}$, thus the total symmetry group should consist of four group elements $\{I,T,P_f\equiv T^2,TP_f\equiv T^3\}$, which is actually a $\mathbb{Z}_4$ group with $T^4=1$. Since $1$D SPT phases are classified by the projective representation of the corresponding symmetry group\cite{XieSPT1,XieSPT2},
the pair of Majorana zero modes $(\gamma_\up,\gamma_\down)$ and $(\gamma_\up^\prime,\gamma_\down^\prime)$ on both ends must carry the $T^4=-1$ projective representation of the bulk $\mathbb{Z}_4$ antiunitary symmetry.

Actually, for the pair of Majorana zero modes $\gamma_{\uparrow}$ and $\gamma_{\downarrow}$ on the left end, we should define its time reversal operator $T_{(\up\down)}$ by $T_{(\up\down)}=U_{(\up\down)}K$, where
$U_{(\up\down)}=\frac{1}{\sqrt{2}}(1+\gamma_\up \gamma_\down)\equiv e^{\frac{\pi}{4}\gamma_\up\gamma_\down}$ and $K$ is the complex conjugate.
$U_{(\up\down)}$ is a unitary operator satisfying
$U_{(\up\down)}U_{(\up\down)}^\dagger=\frac{1}{2}(1+\gamma_\up \gamma_\down)(1-\gamma_\up \gamma_\down)=1$.
Such a definition gives rise to the correct transformation law for $\gamma_\up$ and $\gamma_{\down}$ in Eq.(\ref{T1}).
However, we note that $T^2_{(\up\down)}=\gamma_\up\gamma_\down\neq -1$ satisfies
$T^4_{(\up\down)}=(\gamma_\up\gamma_\down)^2=-1$.
Similarly, for the pair of Majorana zero modes $(\gamma_{\uparrow}^\prime,\gamma_{\downarrow}^\prime)$ on the right end, $T_{(\up^\prime\down^\prime)}$ can be defined by $T_{(\up^\prime\down^\prime)}=U_{(\up^\prime\down^\prime)} K$ with $U_{(\up^\prime\down^\prime)} =\frac{1}{\sqrt{2}}(1+\gamma_\up^\prime \gamma_\down^\prime)=e^{\frac{\pi}{4}\gamma_\up^\prime\gamma_\down^\prime}$.
By using the Fock basis of (local) complex fermion operators $c_L$ and $c_R$ on both ends with
$c_L=\frac{1}{2}(\gamma_\up+i\gamma_\down)$ and $c_R=\frac{1}{2}(\gamma_\up^\prime-i\gamma_\down^\prime)$,
we can derive the representation theory for $T^4=-1$ time reversal symmetry. The higher dimensional generalization for Majorana zero modes realizing $T^4=-1$ time reversal symmetry in point like defects(vortex in $2$D and hedgehog in $3$D) of DIII TSC is straightforward, see supplementary material for details.

{\it $P^4=-1$ parity symmetry, $\overline C^4=-1$ charge conjugation symmetry for a pair of Majorana zero modes} --- Interestingly, we find that a pair of Majorana zero modes can also realize a $P^4=-1$ parity symmetry. For example, in the above $1$D TSC, the parity operator for the pair of Majorana zero modes $(\gamma_\up,\gamma_\up^\prime)$ can be defined by $P_{(\up\up^\prime)}=\frac{1}{\sqrt{2}}(1+\gamma_\up\gamma_\up^\prime )=e^{\frac{\pi}{4}\gamma_\up\gamma_\up^\prime}$. It satisfies $P_{(\up\up^\prime)}^4=-1$ and maps the Majorana mode $\gamma_\up$ on the left end to $\gamma_\up^\prime$ on the right end without spin flipping, with
 $P_{(\up\up^\prime)} \gamma_{\uparrow} P_{(\up\up^\prime)}^{-1}=
-\gamma_{\uparrow}^\prime$ and  $P_{(\up\up^\prime)} \gamma_{\uparrow}^\prime P_{(\up\up^\prime)}^{-1}=
\gamma_{\uparrow}$.
Similarly, for the pair of Majorana zero modes $(\gamma_\down,\gamma_\down^\prime)$, its parity operator can be defined by $P_{(\down \down^\prime)}=\frac{1}{\sqrt{2}}(1-\gamma_\down \gamma_\down^\prime)=e^{-\frac{\pi}{4}\gamma_\down\gamma_\down^\prime}$, which acts on $(\gamma_\down,\gamma_\down^\prime)$ as
$P_{(\down \down^\prime)} \gamma_{\downarrow} P_{(\down \down^\prime)}^{-1} = \gamma_{\downarrow}^\prime$ and
$P_{(\down \down^\prime)} \gamma_{\downarrow}^\prime P_{(\down \down^\prime)}^{-1}= -\gamma_{\downarrow}$.
In additional to the $T^4=-1$/$P^4=-1$ time reversal/parity symmetry, we can even define a $\overline C^4=-1$ nontrivial "charge conjugation" symmetry for a pair Majorana modes $(\gamma_\up,\gamma_\down^\prime)$ or $(\gamma_\down,\gamma_\up^\prime)$. For example, we can define $\overline C_{\up\down^\prime}=\frac{1}{\sqrt{2}}(1+\gamma_\up \gamma_\down^\prime)=e^{\frac{\pi}{4}\gamma_\up\gamma_\down^\prime}$ for the pair of Majorana zero modes $(\gamma_\up,\gamma_\down^\prime)$, and we can define $\overline C_{\down\up^\prime}=\frac{1}{\sqrt{2}}(1+\gamma_\down \gamma_\up^\prime)=e^{\frac{\pi}{4}\gamma_\down\gamma_\up^\prime}$ for the pair of Majorana zero modes $(\gamma_\down,\gamma_\up^\prime)$.

{\it Super $\overline CPT$ algebra for a Majorana fermion} --- The above results from a condensed matter model motivate us to revisit the $\overline C P T$ symmetries\footnote{Since the usual charge conjugation symmetry $C$ has a trivial action on a Majorana fermion(as it is real), here we use the notation $\overline C$ to represent the nontrivial "charge conjugation" symmetry for a Majorana fermion made up of four Majorana zero modes at cutoff.} for a Majorana fermion---a four component real Lorentz spinor, by assuming it is made up of four Majorana zero modes at cutoff energy scale, e.g., a tiny open string described by the 1D TSC in Fig. \ref{chain}.
It is easy to check that the closed algebra of $\overline C,P,T$ symmetries for a Majorana fermion made up of four Majorana zero modes $(\gamma_{\up},\gamma_{\down},\gamma_{\up}^\prime,\gamma_{\down}^\prime)$ satisfies:
\begin{eqnarray}
\overline C^2&=&P^f;\quad P^2=P^f ;\quad T^2=P^f;\quad {(P^f)}^2=1\nonumber\\
TP^f&=&P^fT;\quad PP^f=P^fP;\quad \overline CP^f=P^f \overline C\nonumber\\
TP &=& P^fPT;\quad T\overline C=P^f\overline CT;\quad P\overline C=P^f \overline CP, \label{sual}
\end{eqnarray}
where $\overline C=\overline C_{\up\down^\prime}\otimes \overline C_{\down\up^\prime}$ , $P=P_{(\up\up^\prime)} \otimes P_{(\down \down^\prime)}$, $T=U_{(\up\down)}\otimes U_{(\up^\prime\down^\prime)}K$ and $P^f=\gamma_\up\gamma_\down \gamma_\up^\prime\gamma_\down^\prime$ are the total $\overline C, P, T$ and fermion parity operators for four Majorana zero modes.
Such a $\overline CPT$ algebra is indeed a super algebra, and it is one of the central results of this paper. It arises from the topological nature of a Majorana fermion.

The implementation of the $\overline CPT$ super algebra into quantum field theory is straightforward. We choose four real gamma matrices:
$\gamma_0=-i\rho_z \otimes \sigma_y$, $\gamma_1=I \otimes \sigma_z$, $\gamma_2=-\rho_y \otimes \sigma_y$ and $\gamma_3=-I \otimes \sigma_x$,
where $\rho$ and $\sigma$ are Pauli matrices and $I$ is the identity matrix. We can define a real $\gamma_5$ by $\gamma_5=\gamma_0\gamma_1\gamma_2\gamma_3=i\rho_x\otimes\sigma_y$.
The four component Majorana field
$\psi(x)= \left(
        \begin{array}{c}
         \xi(x)\\
        \eta(x) \\
        \end{array}
      \right)$
can be constructed by two $SO(2)$ real spinor basis
$\xi(x)= \left(
        \begin{array}{c}
          \gamma_\up(x)  \\
          \gamma_\down(x)  \\
        \end{array}
      \right)$ and
$\eta(x)= \left(
        \begin{array}{c}
          -\gamma_\up^\prime(x)  \\
          \gamma_\down^\prime(x) \\
        \end{array}
      \right)$,
which are equivalent to the complex fermion fields $c_L(x)$ and $c_R(x)$. Here $ x=(t,\v x)$ is the four coordinates. The (equal time) canonical commutation relation reads
$\{\psi^\dagger (\v x),\psi (\v y)\}=2\delta^{(3)}(\v x-\v y)$.
In terms of $\gamma_{\sigma}(\v x)$ and $\gamma_{\sigma}^\prime(\v x)$, we have
$\{\gamma_{\sigma}(\v x),\gamma_{\sigma^\prime}^\prime(\v y)\}=0$ and  $\{\gamma_{\sigma}(\v x),\gamma_{\sigma^\prime}(\v y)\}=2\delta^{(3)}(\v x-\v y)\delta_{\sigma\sigma^\prime}$.
The $\overline C,P,T$ symmetry operators and total fermion parity operator can be defined by:
\begin{eqnarray}
\overline C&=&e^{\frac{\pi}{4}\int d^3 x\gamma_{\up}(\v x)\gamma_{\down}^\prime(\v x)}e^{\frac{\pi}{4}\int d^3 x\gamma_{\down}(\v x)\gamma_{\up}^\prime
(\v x)}\nonumber\\
P&=&e^{\frac{\pi}{4}\int d^3x \gamma_{\up}(\v x)\gamma_{\up}^\prime(\v x)}e^{-\frac{\pi}{4}\int d^3 x\gamma_{\down}(\v x)\gamma_{\down}^\prime
(\v x)}P_0\nonumber\\
T&=&e^{\frac{\pi}{4}\int d^3 x\gamma_{\up}(\v x)\gamma_{\down}(\v x)}e^{\frac{\pi}{4}\int d^3 x\gamma_{\up}^\prime(\v x)\gamma_{\down}^\prime
(\v x)}K\nonumber\\
P^f&=&\overline C^2=P^2=T^2
\end{eqnarray}
Here $P_0$ is the action on the spacial coordinates with $P_0 \v x P_0^{-1}=-\v x$. It is easy to check that the above $\overline C,P,T$ symmetry operators satisfy the super algebra Eq.(\ref{sual}).

Under the above $\overline C, P, T$ symmetries, the four component Majorana field $\psi(x)$ transforms as:
\begin{eqnarray}
\overline C \psi(x) \overline C^{-1}&=& \left(
        \begin{array}{c}
         -\epsilon\eta(x)\\
        -\epsilon\xi(x) \\
        \end{array}
      \right)
= -\gamma_5\psi( x);\nonumber\\ P
\psi(x) P^{-1}&=& \left(
        \begin{array}{c}
         \eta(\t x)\\
         -\xi(\t x) \\
        \end{array}
      \right)=\gamma_0\gamma_5 \psi(\t x);\nonumber\\ T \psi(x) T^{-1}&=&\left(
        \begin{array}{c}
         -\epsilon\xi(-\t x)\\
         \epsilon\eta(-\t x) \\
        \end{array}
      \right)=\gamma_0 \psi(-\t x),\label{CPT}
\end{eqnarray}
where $\t x=(t,-\v x)$. Apparently, in the massless limit, the Lagrangian $\mathcal{L}_0=\frac{1}{4}\overline\psi(x)i\gamma_\mu \partial_\mu \psi_c(x)$ with $\overline\psi(x)=\psi^\dagger(x)\gamma_0$ is invariant under the $\overline C,P,T$ symmetries. However, the Majorana mass term $
H_m=\frac{m}{2}\int d^3x \left[i \gamma_\up(\v x) \gamma_{\up}^\prime(\v x)-i \gamma_\down(\v x) \gamma_{\down}^\prime(\v x)\right]$ breaks the charge conjugation symmetry since $\overline C H_m \overline C^{-1}=-H_m$. Such an observation motivates us to elevate the charge conjugation symmetry to a $\mathbb{Z}_2$ gauge symmetry, thus the origin of the Majorana mass term can be explained as the spontaneous gauge symmetry breaking through the Anderson-Higgs mechanism. We introduce a real scalar field $\phi(x)$ which carries $\mathbb{Z}_2$ gauge charge one(thus it transforms as $\overline C \phi(x) \overline C^{-1}=-\phi(x)$) and couple it to the Majorana field. We further assume that such a fundamental scalar field does not carry any other gauge charges and it is invariant under the $P$ and $T$ symmetries.
Finally, we obtain the following Lorentz invariant Lagrangian preserving all the $\overline C,P,T$ symmetries:
\begin{eqnarray}
\mathcal{L}=\frac{1}{4}\overline\psi(x)i\gamma_\mu D_\mu \psi(x)+\frac{ig}{4}\phi(x)\overline\psi(x)\gamma_5\psi(x)+\mathcal{L}_{\phi}+\mathcal{L}_{\mathbb{Z}_2}
\end{eqnarray}
If the real scalar field condenses at $\langle\phi(x)\rangle=\phi_0$, a mass term $im\overline\psi(x)\gamma_5\psi_(x)$ arises with $m=g\phi_0/4$. Here $D_\mu$ represents the covariant derivative, $\mathcal{L}_{\phi}$ is the action of the scale field $\phi(x)$ and $\mathcal{L}_{\mathbb{Z}_2}$ is the action of the $\mathbb{Z}_2$ gauge field.\footnote{We need to regulate the field theory in a discrete space-time to write down its explicit form and we will leave these details in our future publications.} To this end, we explain the origin of Majorana mass for right-handed sterile neutrinos.

\begin{figure}[t]
\begin{center}
\includegraphics[width=6cm]{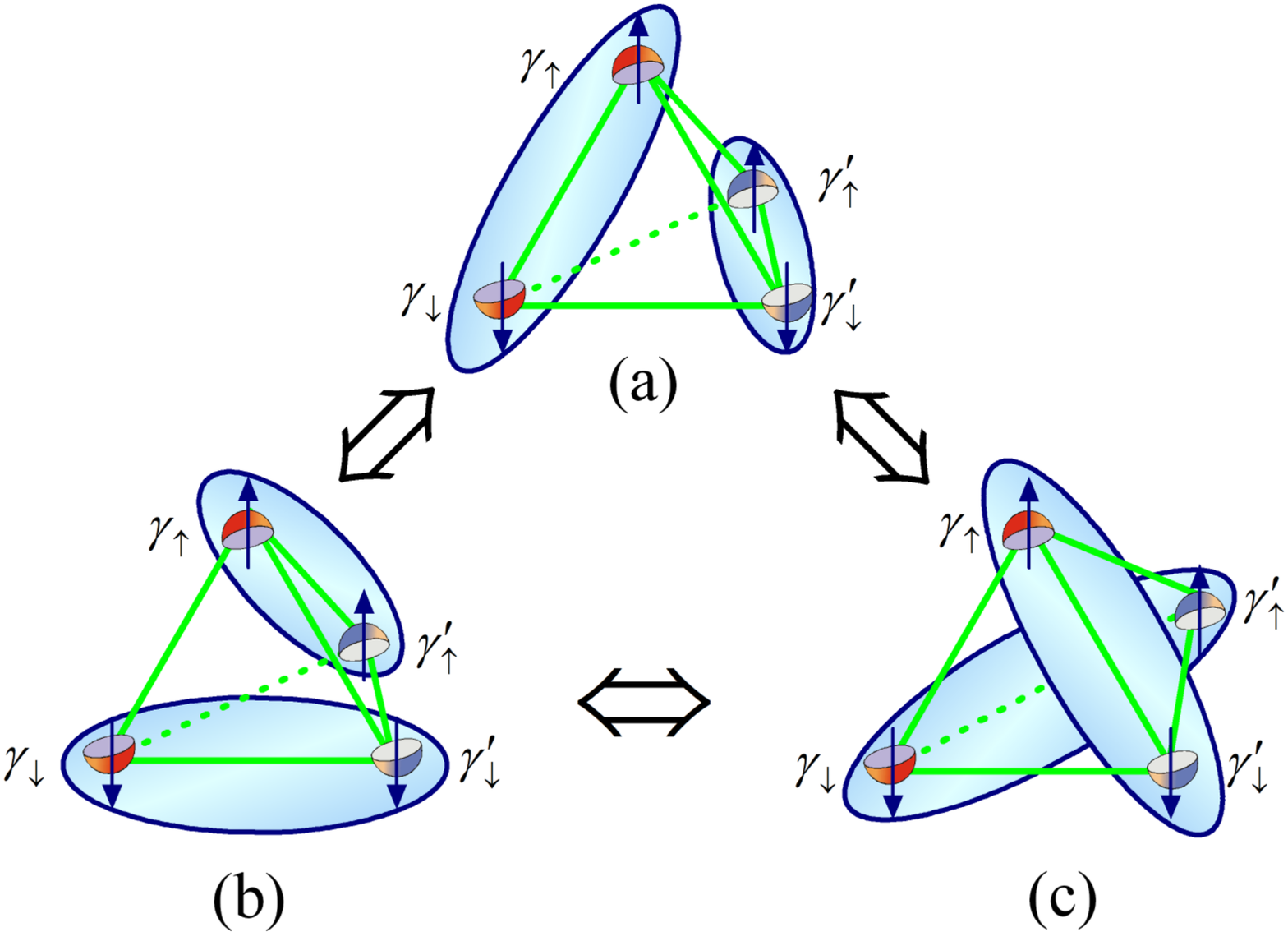}
\caption{(color online)The internal structure of a Majorana fermion indicates that the three generations of neutrinos can be explained by three different ways to form a pair of complex fermions out of four Majorana modes at cutoff scale.
}\label{internal}
\end{center}
\vskip -0.8cm
\end{figure}

{\it The origin of three generations of neutrinos and their mass mixing} --- Surprisingly, the concept that a Majorana fermion is made up of four Majorana zero modes even leads to a natural explanation for the origin of three generations of neutrinos. As we know, due to the Witten anomaly\cite{Wittenanormaly,3DTSC}, a single (local) Majorana zero mode is prohibited in $3D$. Therefore, the four Majorana zero modes must be paired up, and there are three different ways to pair them up. As a pair of Majorana modes is equivalent to a complex fermion mode, we can use the Fock space of three different complex fermions $c_L/c_R$, $d_L=\frac{1}{2}(\gamma_\up-i\gamma_\down^\prime)/d_R=\frac{1}{2}(\gamma_\up^\prime-i\gamma_\down)$ or $f_L=\frac{1}{2}(\gamma_\up+i\gamma_\up^\prime)/f_R=\frac{1}{2}(\gamma_\down+i\gamma_\down^\prime)$ to define the local Hilbert space at cutoff scale, as seen in Fig. \ref{internal}. Furthermore $c_{L(R)}$, $d_{L(R)}$ and $f_{L(R)}$ fermions can be characterized by the $T^4=-1$, $(TP)^4=-1$ and $(T\overline C)^4=-1$ projective representation of $T$, $(TP)$ and $(T\overline C)$ symmetries that they carry. In supplementary material, we construct an explicit lattice model whose local Hilbert space consists of Fock spaces of $c_L$ or $c_R$ fermions, however, its low energy effective theory is a relativistic theory described by Weyl fermions(equivalent to Majorana fermions).

In quantum field theory, the above argument can be implemented by choosing different $SO(2)$ real spinor basis. Apparently, the Majorana field $\psi_c(x)\equiv\psi(x)$ corresponds to $c_L$ and $c_R$ fermions. For $d_L$ and $d_R$ fermions, we define $\psi_d(x)= \left(
\begin{array}{c}
   \hat\xi(x)\\
       \hat\eta(x) \\
\end{array}
\right)$ with $\hat\xi(x)= \left(
        \begin{array}{c}
          \gamma_\up(x)  \\
          \gamma_\down^\prime(x)  \\
        \end{array}
      \right)$ and
$\hat\eta(x)= \left(
        \begin{array}{c}
          \gamma_\down(x)  \\
          -\gamma_\up^\prime(x) \\
        \end{array}
      \right)$, while for $f_L$ and $f_R$ fermions, we define $\psi_f(x)= \left(
\begin{array}{c}
   \t\xi(x)\\
       \t\eta(x) \\
\end{array}
\right) $  with
$\t\eta(x)= \left(
        \begin{array}{c}
          \gamma_\down(x) \\
          \gamma_\down^\prime(x)  \\
        \end{array}
      \right)$
 and
$\t\xi(x)= \left(
        \begin{array}{c}
          \gamma_\up(x)  \\
          \gamma_\up^\prime(x)  \\
        \end{array}
      \right)$.
We note that the Majorana fields $\psi_c(x)$, $\psi_d(x)$ and $\psi_f(x)$ transform differently under $\overline C,P,T$ symmetries(see supplementary material for details), which allow us to compute the neutrino mass mixing matrix from a first principle.

In the extended SM, the total mass matrix has the form $M_{total}=\left(
   \begin{array}{cc}
     0 & m_D \\
     m_D & M \\
   \end{array}
 \right)$,
where $m_D$ is the 3 by 3 Dirac mass matrix and $M$ is the 3 by 3 Majorana mass matrix of three generations of right-handed sterile neutrinos.(We note that the left-handed neutrinos have a zero mass.)With a proper choice of basis, $m_D$ can be chosen to be a diagonal matrix.
We further assume that $m_D$ is uniform,
with the form $m_D=diag(m,m,m)$. The reason why we choose such a form is that the three generations of left-handed/right-handed neutrinos are made up of the same four Majorana zero modes at cutoff scale.
According to the $\mathbb{Z}_2$ gauge (minimal coupling) principle, we can write down the most general $\overline C,P,T$ invariant mass term for three generations of right-handed neutrinos.
\begin{eqnarray}
\mathcal{L}_m &=&\frac{ig}{4}\phi(x)\left[\overline {\psi}_f(x)\psi_f(x)+\overline{\psi}_c(x)\gamma_5\psi_c(x)+\overline{\psi}_d(x)\psi_d(x)\right]
\nonumber\\&+&\frac{ig^\prime}{4}\phi(x)\overline{\psi}_f(x)(1+\gamma_5)\psi_c(x)
+h.c.\nonumber\\&+&\frac{ig^\prime}{4}\phi(x)\overline{\psi}_f(x)(1-\gamma_0\gamma_5)\psi_d(x)
+h.c.\nonumber\\&+&
\frac{ig^\prime}{4}\phi(x)\overline {\psi}_c(x)(1+\gamma_5)(1-\gamma_0\gamma_5)\psi_d(x)+h.c.,
\label{mass}
\end{eqnarray}
where $\overline {\psi}_f(x)={\psi}^\dagger_f(x) \gamma_0$ and $\overline{\psi}_d(x)=\psi^\dagger_d(x) \gamma_5$.
Here we use the same coupling $g$ for all the diagonal mass terms and $g^\prime$ for all the off-diagonal mass terms. Again, this is because the three generations of right-handed neutrinos are made up of the \emph{same} four Majorana zero modes at cutoff scale. \footnote{The above argument can also be incorporated
into traditional quantum field theory language(in the absence of cutoff physics) by imposing a $\mathbb{Z}_2\otimes \mathbb{Z}_2$ flavor symmetry to constraint the coupling constant, see supplementary material for details.}   We note that for $\psi_c(x)$ and $\psi_f(x)$, the boost generators are defined by $S_{0i}=\frac{1}{4}[\gamma_0,\gamma_i]$ while for $\psi_d(x)$, the boost generator is defined by $\bar S_{0i}=\frac{1}{4}[\gamma_5,\gamma_i]$. Such a twisted definition makes the above mass term invariant under the Lorentz transformation.

Since in the extended SM, three generations of right-handed neutrinos are described by three copies of the same Majorana field. Let us redefine $\psi_c(x)$ and $\psi_d(x)$ by $\psi_{c}^\prime(x)\equiv\frac{1+\gamma_5}{\sqrt{2}}\psi_{c}(x)$ and $\psi_d^\prime(x)\equiv\frac{1}{\sqrt{2}}(1-\gamma_0\gamma_5)\psi_d(x)$.
Thus, $\psi_f(x)$, $\psi_{c}^\prime(x)$ and $\psi_d^\prime(x)$ transform in the same way under the $\overline C,P,T$ symmetries. The $\overline C,P,T$ invariant mass term Eq.(\ref{mass}) can be rewritten as:
\begin{eqnarray}
\mathcal{L}_m &=&\frac{ig}{4}\phi(x)\left[\overline {\psi}_f(x)\psi_f(x)+\overline{\psi^\prime}_c(x)\psi^\prime_c(x)+\overline{\psi^\prime}_d (x)\psi_d^\prime(x)\right]
\nonumber\\&+&\frac{\sqrt{2}i g^\prime}{4}\phi(x)\left[\overline{\psi}_f(x)\psi_c^\prime(x)
+\overline{\psi^\prime}_c(x)\psi_f(x)\right]\nonumber\\&+&\frac{\sqrt{2}ig^\prime}{4}\phi(x)\left[\overline{\psi}_f(x){\psi^\prime}_d(x)
+\overline{\psi^\prime}_d(x)\psi_f(x)\right]\nonumber\\&+&\frac{2ig^\prime}{4}\phi(x)\left[\overline {\psi^\prime}_c(x)\psi_d^\prime(x)+\overline {\psi^\prime}_d(x)\psi_c^\prime(x)\right]
\end{eqnarray}
with $\overline\psi^\prime_{c}(x)=(\psi^\prime_{c})^\dagger(x)\gamma_0$ and $\overline\psi^\prime_{d}(x)=(\psi^\prime_{d})^\dagger(x)\gamma_0$.
The mass matrix can be diagonalized by(the basis is ordered as $\psi_f,\psi_c^\prime,\psi_d^\prime$ and $\frac{\phi_0}{4}\equiv\frac{\langle\phi(x) \rangle}{4}$ is set to be $1$):
\begin{eqnarray}
M=\left(
    \begin{array}{ccc}
      g &  \sqrt{2}g^\prime &  \sqrt{2}g^\prime \\
      \sqrt{2}g^\prime & g &  2g^\prime \\
       \sqrt{2}g^\prime &  2g^\prime & g \\
    \end{array}
  \right)=U \left(
                      \begin{array}{ccc}
                        M_1 & 0 & 0 \\
                        0 & M_2 & 0 \\
                        0 & 0 & M_3 \\
                      \end{array}
                    \right)U^T, \label{diag}
\end{eqnarray}
where $M_1=(1-\sqrt{5})g^\prime+g$, $M_2=(1+\sqrt{5})g^\prime+g$, $M_3=-2g^\prime+g$ and
\begin{eqnarray}
U=\left(
                     \begin{array}{ccc}
                                             \sqrt{\frac{5+\sqrt{5}}{10}} &    \sqrt{\frac{5-\sqrt{5}}{10}} &0 \\
                     -\sqrt{\frac{5-\sqrt{5}}{20}} & \sqrt{\frac{5+\sqrt{5}}{20}} & -\frac{1}{\sqrt{2}} \\
                       -\sqrt{\frac{5-\sqrt{5}}{20}}& \sqrt{\frac{5+\sqrt{5}}{20}} & \frac{1}{\sqrt{2}} \\
                      \end{array}
                    \right)
                     \label{mixing}
\end{eqnarray}
In terms of mixing angles, we have $\theta_{12}=31.7^\circ(\tan\theta_{12}=\frac{\sqrt{5}-1}{2})$, $\theta_{23}=-45^\circ$ and $\theta_{13}=0$.
According to the seesaw mechanism, the mixing angles for left-handed light neutrinos take the same form as Eq.(\ref{mixing}), and the minus sign in front of $\theta_{23}$ can be eliminated by a proper gauge choice of the charged lepton basis. The obtained mixing angles, consistent with the golden ratio(GR) pattern that has been proposed phemomelogically\cite{masssymmetry1,masssymmetry2,masssymmetry3,masssymmetry4}, are intrinsically close to the current experimental observations.

Finally, we further argue that the diagonal Yukawa coupling has the same strength as the off-diagonal coupling with $|g|=|g^\prime|$(see supplementary material for details).
The solution $g=-g^\prime$($g>0$) implies $M_1=-M_2=\sqrt{5}g$ and $M_3=3g$, which leads to $m_1/m_3=m_2/m_3=3/\sqrt{5}$ in the limit $m_D\ll M$(here $m_1,m_2$ and $m_3$ are eigen masses of the left-handed light neutrinos) and is allowed by the current experimental observations.\footnote{If we assume that the small mass splitting $\Delta m_{12}$ is due to $CP$ violation and is negligible within LO approximation.} However, the solution $g=g^\prime$ leads to $m_2<m_3<m_1$ which contradicts to the current experimental results of either $m_1\simeq m_2<m_3$(normal hierarchy) or $m_1\simeq m_2>m_3$(inverted hierarchy). Therefore, we choose $g=-g^\prime$ here.
Based on the current experimental data $\Delta m^2_{23}\simeq 2.5\times10^{-3}eV^2$, we obtain $m_1=m_2\simeq0.075eV$ and $m_3\simeq0.056eV$.
Within LO approximation, the neutrino mass mixing matrix takes the form $U^{PMNS}=U\cdot diag(1,e^{\pm i\pi/2},1)$(We note that the Majorana phase $e^{\pm i\pi/2}$ arises from the negative eigenvalue $M_2$). We futher predict the effective mass in neutrinoless double beta decay with  $m_{\beta\beta}\equiv|\sum_i m_{i}{(U_{ei}^{PMNS})}^2|=m_1/\sqrt{5}\simeq 0.034eV$.

{\it Conclusions and discussions} --- In conclusion, based on the assumption that a Majorana fermion is made up of four Majorana zero modes at cutoff, we find a super $\overline C,P,T$ algebra for the Majorana fermion field. Such a concept even successfully explains the origin of three generations of neutrinos and their mass mixing matrix.
Although the obtained mixing angles are as same as the RG pattern, we emphasize that our neutrino mass matrix Eq.(\ref{diag}) actually has an enhanced $D_4$ symmetry(see supplementary materials for details) instead of the standard $\mathbb{Z}_2\otimes\mathbb{ Z}_2$ Klein symmetry\cite{masssymmetry2}. Importantly, our theory not only explains the origin of the discrete flavor symmetry for the neutrino mass matrix, but also completely determines the mass ratios of three generations of neutrinos within LO approximation. (We note that the discrete flavor symmetry is not enough to constraint the neutrino mass ratios.) We predict the mass ratios of three generations to be $m_1/m_3=m_2/m_3=3/\sqrt{5}$ and the effective mass in neutrinoless double beta decay to be $m_{\beta\beta}=m_1/\sqrt{5}$.

Finally, our theory suggests that the nonzero $\Delta m_{12}$ and $\theta_{13}$ have a common origin -- the $CP$ violation correction. This is because the GR pattern has a zero $\theta_{13}$, and if we ignore the charged lepton contribution for $\theta_{13}$ due to its huge mass hierarchy, the experimentally observed nonzero $\theta_{13}$ must come from $CP$ violation correction. On the other hand, as our theory predicts $m_1=m_2$ within LO approximation, the experimentally observed small mass splitting $\Delta m_{12}$ is also contributed by $CP$ violation correction. Interestingly, the current experimental results point to the relation $|\Delta m_{12}/\Delta m_{23}|\sim \theta_{13}/\theta_{23}$.

{\it Acknowledgement} ---
Z-C Gu thanks T.K. Ng for his invitation for IAS Program on
 Topological Materials and Strongly Correlated Electronic Systems at HKUST, where the work was initiated, and thanks Henry Tye, T. Liu for helpful discussions on early results. Z-C Gu thanks John Preskill, Alexei Kitaev, X-G Wen, Y-S Wu, and D. Gaiotto for encouragement and insightful discussions, and especially thanks his wife Y-F Ge for help investigating experimental results. This work is supported by the Government of Canada
through Industry Canada and by the Province of Ontario through the Ministry of
Research and Innovation.

\begin{widetext}
\section{Supplementary material}
\subsection{Representation theory of the $T^4=-1$ time reversal symmetry}
In this section, we work out the explicit representation theory for the $T^4=-1$ time reversal symmetry.
We note that the two pairs of Majorana zero modes on both ends allow us to define two complex fermions $c_L$ and $c_R$:
\begin{eqnarray}
c_L=\frac{1}{2}(\gamma_\up+i\gamma_\down);\quad c_R=\frac{1}{2}(\gamma_\up^\prime-i\gamma_\down^\prime)\label{cLR}
\end{eqnarray}
where $c_L$ and $c_R$ transform nontrivially under the $T^4=-1$ symmetry. We have:
\begin{eqnarray}
T c_{L} T^{-1}&=&- i c_{L}^\dagger;\quad T c_{R} T^{-1}= i c_{R}^\dagger\nonumber\\ T c_{L}^\dagger T^{-1}&=& i c_{L};\quad T c_{R}^\dagger T^{-1}=- i c_{R}\label{TLR}
\end{eqnarray}
Since the $T$ operator only involves two Majorana operators, we are able to construct a representation theory for the $T^4=-1$ symmetry in a two dimensional Hilbert space expanded by two Majorana operators. On the other hand, a projective representation can not be one dimensional, hence we must have:
\begin{eqnarray}
T|\t 0\rangle=UK|\t 0\rangle=U|\t 0\rangle=|\t 1\rangle\equiv c_{L(R)}^\dagger |\t 0\rangle \label{Rep1}
\end{eqnarray}
where $|\t 0\rangle$ is the vacuum of $c_{L(R)}$ fermion satisfying $c_{L(R)}|\t 0\rangle=0$ and $|\t1\rangle\equiv c_{L(R)}^\dagger |\t 0\rangle$. We also assume that the global phase of $|\t 0\rangle$ is fixed in such a way that the complex conjugate $K$ has a trivial action on it. From the relation Eq.(\ref{TLR}), it is straightforward to derive:
\begin{eqnarray}
T |\t 1\rangle &=&UK c_{L(R)}^\dagger |\t 0\rangle=U c_{L(R)}^\dagger |\t 0\rangle\nonumber\\
&=&T c_{L(R)}^\dagger T^{-1} T |\t 0\rangle=\pm ic_{L(R)}c_{L(R)}^\dagger |\t 0\rangle=\pm i|\t 0\rangle \label{Rep2}
\end{eqnarray}
Here the $+$ sign corresponds to $c_L$ and the $-$ sign corresponds to $c_R$.
Thus, in the basis $|\t 0\rangle$ and $|\t 1\rangle$, we can derive the representation theory $T=UK$ with:
\begin{eqnarray}
U=\left(
    \begin{array}{cc}
      0 & 1 \\
      \pm i & 0 \\
    \end{array}
  \right),\label{RepT}
\end{eqnarray}
Clearly, the above representation satisfies $T^4=-1$.

\subsection{ Majorana zero modes in higher dimensions and emergent relativistic dispersion, $SU(2)$ spin}
Majorana zero modes exist in point like defects of $\rm{DIII}$ class TSC in higher dimensions as well. In $2$D, it is well known that a single Majorana zero mode can emerge in the vortex core of a $p+ip$ or $p-ip$ TSC\cite{ReadMajorana}, however, the time reversal symmetry is broken in this class of \emph{chiral} TSC. Nevertheless, the $\rm{DIII}$ class TSC in $2$D that is realized as a composition of a $p+ip$ and a $p-ip$ TSC with opposite spins\cite{2DTSC} can preserve the $T^2=-1$ time reversal symmetry. Apparently, the vortex core of such a TSC has a pair of Majorana zero modes $\gamma_\up$ and $\gamma_\down$ with opposite spins. In the following, we argue that they also carry a $T^4=-1$ representation of time reversal symmetry. As having been discussed in Ref. \cite{2DTSC}, a time reversal action on a single vortex core will change the \emph{local} fermion parity of the complex fermion zero mode $c_L=\gamma_\up+i\gamma_\down$ for the ground state wavefunction, therefore we expect the same representation theory Eq.(\ref{Rep1}),Eq.(\ref{Rep2}) and Eq.(\ref{RepT}) for the zero modes inside the vortex core, which satisfies $T^4=-1$. For the anti-vortex core with Majorana modes
$\gamma_\up^\prime$ and $\gamma_\down^\prime$, we can define a complex fermion mode $c_R=\gamma_\up-i\gamma_\down$ and derive the $T^4=-1$ representation theory as well. Now we see that the $c_L$/$c_R$ complex fermion is similar to the two complex fermion modes defined on the left/right end of the $1$D $T^2=-1$ TSC. The $T^4=-1$ time reversal operators for the Majorana spinons $(\gamma_\up,\gamma_\down)$ and $(\gamma_\up^\prime,\gamma_\down^\prime)$ can be defined in the same way as in $1$D.

The $3$D analogy of the vortex would be a hedgehog and the possibility of the emergence of a Majorana zero mode on the hedgehog has been proposed recently\cite{KaneMajorana}. However, there is an important difference in $3$D. Since the classical configuration of a hedgehog will have a divergent energy, the only way to introduce a UV cutoff is to couple the system to a gauge field, e.g., an $SU(2)$ gauge field\cite{3DTSC}. By turning on the $SU(2)$ gauge field, a single Majorana mode will suffer from the Witten anormally\cite{Wittenanormaly} and the only way to cancel this anormally is to introduce a pair of Majorana zero modes. Therefore, the Majorana zero modes are unstable in the absence of time reversal symmetry(a mass term can be dynamically generated) and the analogy of $p+ip$ TSC does not exist in $3$D. However, in the presence of $T^2=-1$ time reversal symmetry, the pair of Majorana zero modes $\gamma_\up$ and $\gamma_\down$ on the hedgehog can be stabilized(similar to the $1$D and $2$D case, the mass term is forbidden by the time reversal symmetry) and we argue that they also carry a $T^4=-1$ time reversal symmetry according to the same reason as in $2$D -- the time reversal action changes the local fermion parity of the (local) complex fermion mode $c_L=\gamma_\up+i\gamma_\down$ for the ground state wavefunction. The $\rm{DIII}$ class TSC in $3$D labeled by odd integers(there is a $\mathbb{Z}$ classification\cite{Kitaevperiod,Ryuperiod} for free fermion system in this case) could be good candidates to realize a pair of Majorana zero modes on its hedgehog/anti-hedgehog. Detailed discussions of these interesting $3$D models are beyond the scope of this paper and will be presented elsewhere. Finally, we point out an important difference for the Majorana zero modes between $1$D and higher dimensions. In $1$D, for a generic Hamiltonian, the zero modes are only well defined in the infinite long chain limit. However, in $2(3)$D, the distance between vortex(hedgehog) and anti-vortex(anti-hedgehog) can be finite(but much larger than penetration depth) since the zero modes are well defined bound states and they can be regarded as \emph{local} particles.

\begin{figure}[t]
\begin{center}
\includegraphics[width=10cm]{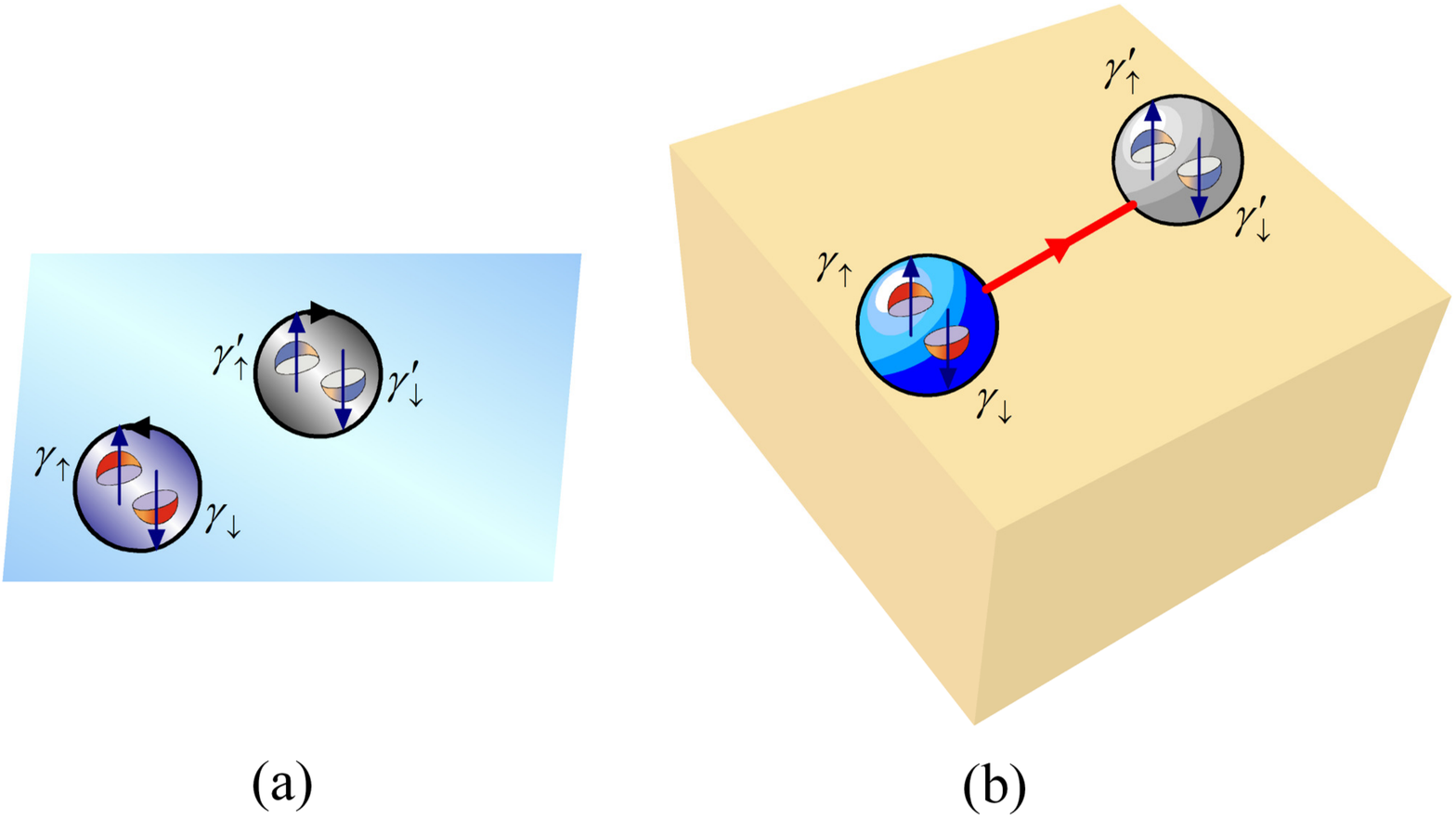}
\caption{(color online)Majorana zero modes in $2$D and $3$D can be realized as the bound states on the vortex/anti-vortex core and hedgehog/anti-hedgehog core of $\rm{DIII}$ class TSC. The red line in (b) represents a quantized flux line that connects a pair of hedgehog and anti-hedgehog.
}\label{hihgerD}
\end{center}
\end{figure}

Now, let us construct a quantum critical model with emergent Majorana fermions in $3$D. First, we construct a $3$D cubic lattice model consisting of hedgehog/anti-hedgehog, with hedgehog occupied sublattice A and anti-hedgehog occupied sublattice B, as seen in Fig \ref{3Dchain}. We use red dots to represent the pair of Majorana modes $(\gamma_\up, \gamma_\down)$ on the hedgehog and blue dots to represent the pair of Majorana modes $(\gamma_\up^\prime, \gamma_\down^\prime)$ on the anti-hedgehog.
Then we turn on the hoping terms among those Majorana modes and consider the following Hamiltonian:
\begin{eqnarray}
H_{3D}&=&-\sum_{\v i\in A; \v j=\v i \pm \hat{\v x}}\left( i\gamma_{\v i,\up} \gamma_{\v j,\down}^\prime +i\gamma_{\v i,\down} \gamma_{\v j,\up}^\prime\right)+\sum_{\v i\in A; \v j=\v i \pm \hat{\v y}}\left( i\gamma_{\v i,\up} \gamma_{\v j,\up}^\prime -i\gamma_{\v i,\down} \gamma_{\v j,\down}^\prime\right)
\nonumber\\&+&\sum_{ \v i \in A; \v j=\v i+ \hat{\v z}}\left( i\gamma_{\v i,\up} \gamma_{\v j,\down} -i\gamma_{\v i,\down} \gamma_{\v j,\up}\right)+\sum_{ \v i \in B; \v j=\v i+ \hat{\v z}}\left(i\gamma_{\v i,\up}^\prime \gamma_{\v j,\down}^\prime  -i\gamma_{\v i,\down}^\prime  \gamma_{\v j,\up}^\prime \right)\label{3DMajorana}
\end{eqnarray}
In terms of complex fermions $c_{\v i, L}=\gamma_{\v i, \up}+i\gamma_{\v i,\down}$ and $c_{\v i, R}=\gamma_{\v i,\up}^\prime-i\gamma_{\v i\down}^\prime$, we have:
\begin{eqnarray}
H_{3D}&=&\sum_{\v i\in A; \v j=\v i \pm \hat{\v x}}\left( c_{L,\v i}^\dagger c_{R,\v j}+c_{R,\v j}^\dagger c_{L,\v i}\right)+i\sum_{\v i\in A; \v j=\v i \pm \hat{\v y}}\left( c_{L,\v i}^\dagger c_{R,\v j}-c_{R,\v j}^\dagger c_{L,\v i}\right)
\nonumber\\&+&\sum_{ \v i \in A; \v j=\v i+ \hat{\v z}}\left( c_{L,\v i}^\dagger c_{L,\v j}+ c_{L,\v j}^\dagger c_{L,\v i}\right)-\sum_{ \v i \in B; \v j=\v i+ \hat{\v z}}\left(c_{R,\v i}^\dagger c_{R,\v j}+c_{R,\v j}^\dagger c_{R,\v i}\right)\label{3Dlattice}
\end{eqnarray}

\begin{figure}[t]
\begin{center}
\includegraphics[width=10cm]{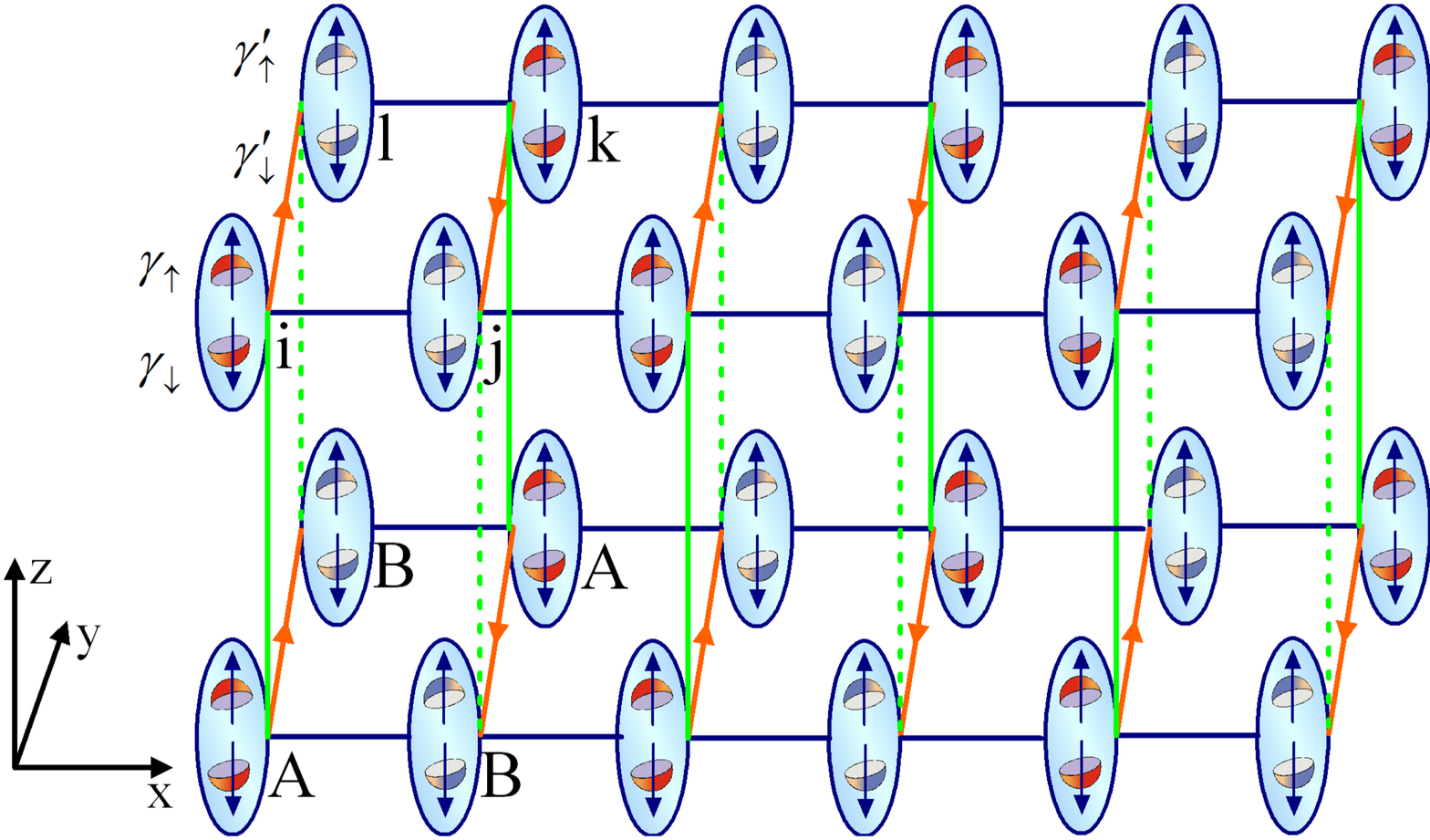}
\caption{(color online)A $3$D cubic lattice model consists of hedgehogs/anti-hedgehogs. Red dots represent the pair of Majorana modes $(\gamma_\up, \gamma_\down)$ on the hedgehog and blue dots represent the pair of Majorana modes $(\gamma_\up^\prime, \gamma_\down^\prime)$ on the anti-hedgehog. Solid/dashed lines represent the hopping amplitude $1/-1$. Lines with arrows represent the hopping amplitudes $\pm i$. Multiplications of the hopping amplitudes surround a square surface give rise to $-1$, e.g., $t_{ij}t_{jk}t_{kl}t_{li}=-1$.
Such a hopping amplitudes pattern is the so called $\pi$-flux pattern.
}\label{3Dchain}
\end{center}
\end{figure}

The special hopping pattern in the above Hamiltonian is one way to realize the so called $\pi$-flux pattern, namely, a pattern with the enclosed flux $\pi$ on each face of the cubic lattice.
The Hamiltonian is invariant under the time reversal symmetry $\t T=T^{(-)^{i_z}}$. (Without such a twisted definition of the time reversal symmetry, the fermion hopping in the $z$-direction will change sign under the time reversal. It is clear that such a twisted definition is allowed because we can choose either $T$ or $T^{-1}$ as the definition of the time reversal symmetry.)

In momentum space, we have:
\begin{eqnarray}
H_{3D}=\sum_{\v k} (c_L^\dagger(\v k),c_R^\dagger(\v k))\left(
                                            \begin{array}{cc}
                                              2\cos k_z & 2\cos k_x+2i\cos k_y \\
                                              2\cos k_x-2i\cos k_y & -2\cos k_z  \\
                                            \end{array}
                                          \right)\left(\begin{array}{c}
                                                         c_L(\v k) \\
                                                         c_R(\v k)
                                                       \end{array}\right)
\end{eqnarray}
The above Hamiltonian has one positive energy mode and one negative energy mode with:
\begin{eqnarray}
E_{\v k}=\pm 2\sqrt{\cos^2 k_x+\cos^2 k_y+\cos^2 k_z},
\end{eqnarray}
Around the momentum point $\v k_0=(\pi/2,\pi/2,\pi/2)$, the above Hamiltonian describes a chiral Weyl fermion(equivalent to a Majorana fermion):
\begin{eqnarray}
H^{eff}_{(\pi/2,\pi/2,\pi/2)}=2\sum_{\v k} (c_L^\dagger(\v k),c_R^\dagger(\v k))\left(
                                            \begin{array}{cc}
                                              \bar k_z & \bar k_x+i\bar k_y \\
                                             \bar k_x-i\bar k_y & -\bar k_z  \\
                                            \end{array}
                                          \right)\left(\begin{array}{c}
                                                         c_L(\v k) \\
                                                         c_R(\v k)
                                                       \end{array}\right)
\end{eqnarray}
where $\v k=\v k_0+ \bar {\v k}$. It is clear that the above Hamiltonian has a relativistic dispersion $E_{\v k}=\pm 2|\bar {\v k}|$ and an emergent $SU(2)$ spin carrying angular momentum.

In the above, we construct a particular $3$D hedgehog/anti-hedgehog lattice model with proliferated Majorana zero modes. Other models with deconfined Majorana modes have also been considered recently, e.g., the fermion dimer model\cite{gapless3DTSC} and the Majorana flat bands model in certain gapless TSC\cite{gaplessMajorana}. However, one of the most important features in our model is that it has a sublattice structure, and the sublattice degeneracy naturally leads to an $SU(2)$ spin degree of freedom at low energy. Actually, our model can be viewed as the $3$D analogy of the $2$D graphene system, where the valley degeneracy becomes the emergent $SU(2)$ spin at low energy. But why hedgehog/anti-hedgehog lattice models with a sublattice structure is more natural than those models without sublattice structure? One possible reason is that the hedgehog and anti-hedgehog pair are always confined in a superconductor\cite{monopoleconfinement}, therefore any stable $3$D hedgehog/anti-hedgehog lattice model must contain a hedgehog and anti-hedgehog pair per unit cell.

Our analysis for condensed matter systems implies that the presence of $SU(2)$ spin at low energy has a deep relationship with the sublattice structure at cutoff scale. A very interesting question is that whether the $SU(2)$ spin for all the fundamental particles arises from a similar discrete structure at cutoff scale. Unfortunately, it is very difficult to examine the above idea theoretically since a quantum field theory with an explicit cutoff is absent so far.
Although lattice models could be thought as a natural venue to regulate the theory, any pre-assuming lattice structure for space-time will break the Lorentz invariance seriously. To overcome this difficulty, a discrete topological non-linear sigma model with a dynamic background is a promising candidate, where the sublattice structures correspond to to the opposite orientations of branched tetrahedron. Important progress along this direction has been made recently\cite{XieSPT3,XieSPT4}, even with super-coordinates\cite{Gusuper}. It would be very interesting to examine these ideas in future.

\subsection{Quantum field theory description for three generations of neutrinos}
The quantum field theory description for the Majorana field made up of $c_{L}$ and $c_{R}$ local complex fermions has already been presented in the main text.
To describe the Majorana field made up of $f_{L}$ and $f_{R}$ local complex fermions in quantum field theory, we just need to define the Majorana fermion field $\psi_f(x)= \left(
\begin{array}{c}
   \t\xi(x)\\
       \t\eta(x) \\
\end{array}
\right) $ with a different $SO(2)$ spinor basis:
\begin{eqnarray}
\t\xi(x)= \left(
        \begin{array}{c}
          \gamma_\up(x)  \\
          \gamma_\up^\prime(x)  \\
        \end{array}
      \right);\quad
\t\eta(x)= \left(
        \begin{array}{c}
          \gamma_\down(x) \\
          \gamma_\down^\prime(x)  \\
        \end{array}
      \right),
\end{eqnarray}
The above Majorana fermion field satisfies the $\overline CPT$ symmetries:
\begin{eqnarray}
 \overline C \psi_f(x) \overline C^{-1}=-\gamma_5\psi_f(x); \quad P  \psi_f(x) P^{-1}= \gamma_0 \psi_f(\t x);\quad T \psi_f(x) T^{-1}=-\gamma_0\gamma_5\psi_f(-\t x),
\end{eqnarray}
It is clear that the $f_{L(R)}$ fermion transforms differently under $\overline CPT$ symmetries, and for the $f_{L(R)}$ fermion, its mass term takes the usual form:
\begin{eqnarray}
\mathcal{L}_m=\frac{ig}{4}\phi(x)\overline\psi_f(x)\psi_f(x),\quad \overline\psi_f(x)= \psi_f^\dagger (x) \gamma_0
\end{eqnarray}

For the Majorana fermion made up of $d_{L}$ and $d_{R}$ local complex fermions, we need to choose $\bar\gamma_0=R\gamma_0 R^{-1}=i\rho_x\otimes\sigma_y\equiv\gamma_5$ with:
\begin{eqnarray}
R=\frac{1}{\sqrt{2}}\left(
        \begin{array}{cc}
          1 & 1 \\
          -1 & 1 \\
        \end{array}
      \right)=\frac{1}{\sqrt{2}}(1+\gamma_0\gamma_5)
\end{eqnarray}
The corresponding $\gamma_{1,2,3}$ and  $\gamma_5$ transform as:
$\bar \gamma_{1,2,3}=R\gamma_{1,2,3}R^{-1}=\gamma_{1,2,3}$ and $\bar\gamma_5=R\gamma_5 R^{-1}=i\rho_z\otimes\sigma_y\equiv-\gamma_0$). Indeed, this representation was first proposed by Ettore Majorana.

The quantum field theory can be obtained by defining $\psi_d(x)= \left(
\begin{array}{c}
   \hat\xi(x)\\
       \hat\eta(x) \\
\end{array}
\right)$ with:
\begin{eqnarray}
\hat\xi(x)= \left(
        \begin{array}{c}
          \gamma_\up(x)  \\
          \gamma_\down^\prime(x)  \\
        \end{array}
      \right);\quad
\hat\eta(x)= \left(
        \begin{array}{c}
          \gamma_\down(x)  \\
          -\gamma_\up^\prime(x) \\
        \end{array}
      \right),
\end{eqnarray}
Under the $\overline C P T$ symmetries with above definition, $\psi_d(x)$ transforms as:
\begin{eqnarray}
 \overline C \psi_d(x)  {\overline C}^{-1}&=&-\bar\gamma_5\psi_d(x)\equiv\gamma_0\psi_d(x); \nonumber\\ P \psi_d(x) {P}^{-1}&=& \bar\gamma_0\psi_d(\t x)\equiv\gamma_5\psi_d(\t x);\nonumber\\
 T \psi_d(x) {T}^{-1}&=&-\bar\gamma_0\bar\gamma_5\psi_d(-\t x)\equiv-\gamma_0\gamma_5\psi_d(-\t x),
\end{eqnarray}
Again, the mass term also takes the usual form:
\begin{eqnarray}
\mathcal{L}_m=\frac{ig}{4}\phi(x)\overline\psi_d(x)\psi_d(x),\quad \overline\psi_d(x)= \psi_d^\dagger (x)\bar \gamma_0=\psi_d^\dagger (x) \gamma_5 \label{dfield}
\end{eqnarray}
Since $\psi_{c}$, $\psi_{f}$ and $\psi_{d}$ transform differently under the $\overline CPT$ symmetries and one can not transform them from one to the other with continuous proper orthochronous Lorentz transformation, they can be regarded as three independent fields in quantum field theory.(We note that any continuous  proper orthochronous Lorentz transformation will not change the definition of $SO(2)$ spinor basis.)

\subsection{$\overline CPT$ symmetries in momentum space}
\label{App:momentum}
In this section, we will use a momentum space picture to describe the three generations of neutrinos. First, let us examine the $\overline CPT$ symmetry transformation of the Fourier modes $\gamma_\sigma(\v k)=\frac{1}{\sqrt{V}}\int d^3 x e^{-i\v k \cdot \v x}\gamma_\sigma(\v x)$ and
$\gamma_\sigma^\prime(\v k)=\frac{1}{\sqrt{V}}\int d^3 x e^{-i\v k \cdot \v x}\gamma_\sigma^\prime(\v x)$. It is straightforward to derive:
\begin{eqnarray}
\overline C \gamma_{\uparrow}(\v k) \overline C^{-1} &=&
-\gamma_{\downarrow}^\prime (\v k); \quad
\overline C \gamma_{\downarrow}(\v k) \overline C^{-1}= -\gamma_{\uparrow}^\prime(\v k); \nonumber\\
\overline C \gamma_{\uparrow}^\prime(\v k) \overline C^{-1} &=&
\gamma_{\downarrow}(\v k); \quad
\overline C \gamma_{\downarrow}^\prime(\v k) \overline C^{-1}= \gamma_{\uparrow}(\v k),
\end{eqnarray}
\begin{eqnarray}
P \gamma_{\uparrow}(\v k) P^{-1}&=&
-\gamma_{\uparrow}^\prime(-\v k); \quad
P \gamma_{\downarrow}(\v k) P^{-1}= \gamma_{\downarrow}^\prime(-\v k);\nonumber\\
P \gamma_{\uparrow}^\prime(\v k) P^{-1}&=&
\gamma_{\uparrow}(-\v k); \quad
P \gamma_{\downarrow}^\prime(\v k) P^{-1}= -\gamma_{\downarrow}(-\v k),
\end{eqnarray}
\begin{eqnarray}
T \gamma_{\uparrow}(\v k) T^{-1} &=&
-\gamma_{\downarrow}(-\v k); \quad
T \gamma_{\downarrow}(\v k) T^{-1}= \gamma_{\uparrow}(-\v k); \nonumber\\
T \gamma_{\uparrow}^\prime(\v k) T^{-1} &=&
-\gamma_{\downarrow}^\prime(-\v k); \quad
T \gamma_{\downarrow}^\prime(\v k) T^{-1}= \gamma_{\uparrow}^\prime(-\v k),
\end{eqnarray}

We can apply the similar argument to the emergence of three generations of Majorana fermions for their Fourier modes in momentum space as well.
\begin{eqnarray}
d_L(\v k)&=& \gamma_\up(\v k)-i \gamma_\down^\prime(\v k);\quad d_R(\v k)=\gamma_\up^\prime(\v k)-i \gamma_\down(\v k)\nonumber\\
c_L(\v k)&=& \gamma_\up(\v k)+i \gamma_\down(\v k);\quad c_R(\v k)= \gamma_\up^\prime(\v k)-i \gamma_\down^\prime(\v k)\nonumber\\
f_L(\v k)&=& \gamma_\up(\v k)+i \gamma_\up^\prime(\v k);\quad f_R(\v k)=\gamma_\down(\v k)+i \gamma_\down^\prime(\v k)
\end{eqnarray}
Under $TP,T$ and $T \overline C$ symmetries, they transform as:
\begin{eqnarray}
(TP) d_L(\v k) (TP)^{-1}&=&-i d_L^\dagger(-\v k); \quad (TP) d_R(\v k) (TP)^{-1}=i d_R^\dagger(-\v k) \nonumber\\
T c_L(\v k) T^{-1}&=&-i c_L^\dagger(\v k); \quad T c_R(\v k) T^{-1}=i c_R^\dagger(\v k) \nonumber\\
(T\overline C) f_L(\v k) (T \overline C)^{-1}&=&-i f_L^\dagger(\v k); \quad (T \overline C) f_R(\v k) (T \overline C)^{-1}=i f_R^\dagger(\v k),
\end{eqnarray}

The Hamiltonian of massless Majorana fermion has the following form in momentum space, e.g., for $\psi_d(x)$:
\begin{eqnarray}
\mathcal{H}_d=\frac{1}{4}\sum_{\v k}\psi^\dagger(\v k) \bar \gamma_0 \bar \gamma_i k_i \psi(\v k),
\end{eqnarray}
where $\psi(\v k)$ is the Fourier mode of $\psi(\v x)$, defined as $\psi(\v k)=\frac{1}{\sqrt{V}}\int d^3 x e^{-i\v k \cdot \v x}\psi(\v x)$. It is straightforward to verify that $\psi^\dagger(\v k)=\psi^t(-\v k)$.
If we assume the chiral basis has a spin polarization in the $y$-direction, we can fix the momentum to be $\v k=(0,k,0)$. Thus, we obtain:
\begin{eqnarray}
\mathcal{H}_d=\frac{1}{4}\sum_{k}\left[k\gamma_\up(-k)\gamma_\up(k)-k\gamma_\down(-k)\gamma_\down(k)-k\gamma_\up^\prime(-k)\gamma_\up^\prime(k)
+k\gamma_\down^\prime(-k)\gamma_\down^\prime(k)\right]
\end{eqnarray}
In terms of the chiral fermion fields $d_L(\v k)$ and $d_R(\v k)$, we have:
\begin{eqnarray}
\mathcal{H}_d=\frac{1}{2}\sum_{k}\left[k d_L^\dagger(k) d_L(k)-k d_R^\dagger(k) d_R(k)\right]
\end{eqnarray}
For any given momentum $\v k$, we can define its positive energy mode as a left-handed neutrino and the negative energy mode as a right-handed antineutrino. However, we note that the zero energy mode $d_L(0)$ and $d_R(0)$ still transform as:
\begin{eqnarray}
(TP) d_L(0) (TP)^{-1}&=&-i d_L^\dagger(0); \quad (TP) d_R(0) (TP)^{-1}=i d_R^\dagger(0),
\end{eqnarray}
Thus, both of them carry the $(TP)^4=-1$ fractionalized symmetry. Furthermore, since the zero energy mode $d_{L(R)}$ transforms trivially under Lorentz symmetry, we can say that the vacuum effectively carries a $(TP)^4=-1$ fractionalized symmetry. Such an observation is pretty interesting, as traditional quantum field theory assume a unique vacuum that carries a trivial representation of $TP$ symmetry. The experimental consequence of such a fractionalized symmetry will be investigated in our future work.

For $c_{L(R)}$ and $f_{L(R)}$, their Hamiltonian in momentum space read:
\begin{eqnarray}
\mathcal{H}_{c(f)}=\frac{1}{4}\sum_{\v k}\psi^\dagger(\v k)  \gamma_0 \gamma_i k_i \psi(\v k),
\end{eqnarray}
If we assume the chiral basis has a spin polarization in the $z$-direction, we can fix the momentum to be $\v k=(0,0,k)$
In terms of the $c_{L(R)}$ and $f_{L(R)}$ fermion operators, we have:
\begin{eqnarray}
\mathcal{H}_d&=&\frac{1}{2}\sum_{k}\left[k c_L^\dagger(k) c_L^\dagger(-k)-k c_R^\dagger(k) c_R^\dagger(-k)+h.c.\right];\nonumber\\
\mathcal{H}_f&=&\frac{1}{2}\sum_{k}\left[k f_L^\dagger(k) f_L^\dagger(-k)-k f_R^\dagger(k) f_R^\dagger(-k)+h.c.\right],
\end{eqnarray}
In the Nambu basis, we obtain:
\begin{eqnarray}
\mathcal{H}_c&=&\frac{1}{2}\sum_{k}\left[\begin{array}{cc}
c_L^\dagger(k)+c_R^\dagger(k), & c_L(-k)-c_R(-k)
\end{array}\right]\left(
\begin{array}{cc}
  0 & k \\
    k & 0 \\
    \end{array}
    \right)\left[\begin{array}{c}
    c_L(k)+c_R(k) \\
    c_L^\dagger(-k)-c_R^\dagger(-k)
    \end{array}\right],\nonumber\\
\mathcal{H}_f&=&\frac{1}{2}\sum_{k}\left[\begin{array}{cc}
f_L^\dagger(k)-f_R^\dagger(k), & f_L(-k)+f_R(-k)
\end{array}\right]\left(
\begin{array}{cc}
  0 & k \\
    k & 0 \\
    \end{array}
    \right)\left[\begin{array}{c}
    f_L(k)-f_R(k) \\
    f_L^\dagger(-k)+f_R^\dagger(-k)
    \end{array}\right],
\end{eqnarray}
After diagonalizing the above two Hamiltonians, we can again define a positive mode corresponding to the left-handed neutrino and a negative energy mode corresponding to the right-handed antineutrino.

Similarly, the zero energy mode $c_{L(R)}(0)$ and $f_{L(R)}(0)$ transform as:
\begin{eqnarray}
T c_L(0) T^{-1}&=&-i c_L^\dagger(0); \quad T c_R(0) T^{-1}=i c_R^\dagger(0),
\end{eqnarray}
and
\begin{eqnarray}
(T \overline C) f_L(0) (T \overline C)^{-1}&=&-i f_L^\dagger(0); \quad (T \overline C) f_R(0) (T \overline C)^{-1}=i f_R^\dagger(0),
\end{eqnarray}
Therefore, we can say that the vacuum for for $c_{L(R)}$ and $f_{L(R)}$ fermions can effectively carry $T^4=-1$ and $(T \overline C)^4=-1$ fractionalized symmetries.

\subsection{The $\mathbb{Z}_2\otimes \mathbb{Z}_2$ flavor symmetry and beyond}
\label{App:symmetry}
In this section, we provide a symmetry argument for the choice of Yukawa couplings. Let us start with the diagonal term and assume there are three independent couplings $g_{f}$, $g_d$ and $g_c$.
\begin{eqnarray}
\mathcal{L}_{m-d} &=&\frac{i}{4}\phi(x)\left[g_f\overline {\psi}_f(x)\psi_f(x)+g_d\overline{\psi}_d(x)\psi_d(x)+g_c\overline{\psi}_c(x)\gamma_5\psi_c(x)\right]\label{massdiag}
\end{eqnarray}
According to the definitions of $\psi_f(x)$ and $\psi_c(x)$, they are related to each other by a $\mathbb{Z}_2$ symmetry transformation $\psi_c(x)=S_1\psi_f(x)$, where:
\begin{eqnarray}
S_1=\left(
      \begin{array}{cccc}
        1 & 0 & 0 & 0 \\
        0 & 0 & 1 & 0 \\
        0 & -1 & 0 & 0 \\
        0 & 0 & 0 & 1 \\
      \end{array}
    \right)
\end{eqnarray}
Let us rewrite the diagonal mass term as:
\begin{eqnarray}
\mathcal{L}_{m-d} &=&\frac{i}{4}\phi(x)\left[g_f \overline\psi_f(x) S_1^{-1}S_1\psi_f(x)+g_d\overline{\psi}_d(x)\psi_d(x)+g_c\overline{\psi}_c(x)\gamma_5S_1S_1^{-1}\psi_c(x)\right]\nonumber\\
&=&\frac{i}{4}\phi(x)\left[g_f \psi_f(x)^\dagger S_1^{-1}S_1\gamma_0   S_1^{-1}S_1\psi_f(x)+g_d\overline{\psi}_d(x)\psi_d(x)+g_c{\psi}_c(x)^\dagger S_1 S_1^{-1}\gamma_0\gamma_5S_1S_1^{-1}\psi_c(x)\right]\nonumber\\
&=&\frac{i}{4}\phi(x)\left[g_f \psi_f(x)^\dagger S_1^{-1}\gamma_0 \gamma_5S_1\psi_f(x)+g_d\overline{\psi}_d(x)\psi_d(x)+g_c{\psi}_c(x)^\dagger S_1 \gamma_0S_1^{-1}\psi_c(x)\right]\nonumber\\
&=&\frac{i}{4}\phi(x)\left[g_f \psi_c(x)^\dagger \gamma_0 \gamma_5\psi_c(x)+g_d\overline{\psi}_d(x)\psi_d(x)+g_c{\psi}_f(x)^\dagger \gamma_0\psi_f(x)\right]\nonumber\\
&=&\frac{i}{4}\phi(x)\left[g_f \overline \psi_c(x) \gamma_0 \gamma_5\psi_c(x)+g_d\overline{\psi}_d(x)\psi_d(x)+g_c \overline {\psi}_f(x) \psi_f(x)\right]\label{massdiag1}
\end{eqnarray}
Comparing Eq.(\ref{massdiag}) and Eq.(\ref{massdiag1}), we obtain $g_c=g_f$.

Similarly, $\psi_f(x)$ and $\psi_d(x)$ are also related by another $\mathbb{Z}_2$ symmetry transformation $\psi_d(x)=S_2\psi_f(x)$ with:
\begin{eqnarray}
S_2=\left(
      \begin{array}{cccc}
        1 & 0 & 0 & 0 \\
        0 & 0 & 0 & 1 \\
        0 & 0 & 1 & 0 \\
        0 & -1 & 0 & 0 \\
      \end{array}
    \right)
\end{eqnarray}
Again, we can rewrite the diagonal mass term as:
\begin{eqnarray}
\mathcal{L}_{m-d} &=&\frac{i}{4}\phi(x)\left[g_f \overline\psi_f(x) S_2^{-1}S_2\psi_f(x)+g_d\overline {\psi}_d(x) S_2S_2^{-1}\psi_d(x)+g_c\overline{\psi}_c(x)\gamma_5\psi_c(x)\right]\nonumber\\
 &=&\frac{i}{4}\phi(x)\left[g_f \psi_f(x)^\dagger \gamma_0 S_2^{-1}S_2\psi_f(x)+g_d{\psi}_d(x)^\dagger \gamma_5 S_2S_2^{-1}\psi_d(x)+g_c\overline{\psi}_c(x)\gamma_5\psi_c(x)\right]\nonumber\\
  &=&\frac{i}{4}\phi(x)\left[g_f \psi_f(x)^\dagger S_2^{-1}S_2 \gamma_0S_2^{-1} S_2\psi_f(x)+g_d{\psi}_d(x)^\dagger S_2S_2^{-1}\gamma_5 S_2S_2^{-1}\psi_d(x)+g_c\overline{\psi}_c(x)\gamma_5\psi_c(x)\right]\nonumber\\
  &=&\frac{i}{4}\phi(x)\left[g_f \psi_f(x)^\dagger S_2^{-1}  \gamma_5S_2\psi_f(x)+g_d{\psi}_d(x)^\dagger S_2\gamma_0 S_2^{-1}\psi_d(x)+g_c\overline{\psi}_c(x)\gamma_5\psi_c(x)\right]\nonumber\\
  &=&\frac{i}{4}\phi(x)\left[g_f \psi_d(x)^\dagger   \gamma_5\psi_d(x)+g_d{\psi}_f(x)^\dagger \gamma_0 \psi_f(x)+g_c\overline{\psi}_c(x)\gamma_5\psi_c(x)\right]\nonumber\\
  &=&\frac{i}{4}\phi(x)\left[g_f \overline\psi_d(x)   \gamma_5\psi_d(x)+g_d\overline{\psi}_f(x) \gamma_0 \psi_f(x)+g_c\overline{\psi}_c(x)\gamma_5\psi_c(x)\right]
 \label{massdiag2}
\end{eqnarray}
Comparing Eq.(\ref{massdiag}) and Eq.(\ref{massdiag2}), we obtain $g_d=g_f$. Finally, we have $g_c=g_d=g_f=g$.
Now we see that in traditional quantum field theory language, the choice of diagonal Yukawa couplings can be achieved by imposing the above two $\mathbb{Z}_2$ symmetries, which leads to a $\mathbb{Z}_2\otimes \mathbb{Z}_2$ flavor symmetry.

Nevertheless, traditional quantum field theory can not tell us why there are three generations of neutrinos and where the $\mathbb{Z}_2\otimes \mathbb{Z}_2$ flavor symmetry comes from. To understand these mysteries, the internal structure proposed in this paper -- a Majorana fermion is made up of four Majorana zero modes plays a crucial role. At cutoff scale, all the mass terms should be regarded as interactions between the scalar particle $\phi$ and the four Majorana modes $\gamma_\up,\gamma_\down,\gamma_\up^\prime$ and $\gamma_\down^\prime$. For example, all the three terms in Eq.(\ref{massdiag}) can be expressed as:
\begin{eqnarray}
\frac{ig_f}{4}\phi(x)\overline {\psi}_f(x)\psi_f(x)&=&\frac{ig_f}{2}\phi(x)\left[\gamma_\up(x)\gamma_\up^\prime(x)-\gamma_\down(x)\gamma_\down^\prime(x)\right];\nonumber\\
\frac{ig_d}{4}\phi(x)\overline{\psi}_d(x)\psi_d(x)&=&
\frac{ig_d}{2}\phi(x)\left[\gamma_\up(x)\gamma_\up^\prime(x)-\gamma_\down(x)\gamma_\down^\prime(x)\right];\nonumber\\
\frac{ig_c}{4}\phi(x)\overline{\psi}_c(x)\gamma_5\psi_c(x)&=&
\frac{ig_c}{2}\phi(x)\left[\gamma_\up(x)\gamma_\up^\prime(x)-\gamma_\down(x)\gamma_\down^\prime(x)\right],
\end{eqnarray}
The above expression implies that the three mass terms are indeed the same term at cutoff scale. In terms of traditional quantum field theory language,
we can attribute the existence of three generations of neutrinos to the three different (local) ways of making a pair of complex fermions out of four Majorana zero modes. Therefore, the $\mathbb{Z}_2\otimes \mathbb{Z}_2$ flavor symmetry is indeed a gauge symmetry from our perspective and we obtain $g_c=g_d=g_f=g$. However, at this point, one may confuse that if the three mass terms are the same, why we observe three generations of neutrinos rather than one.  This is because discrete gauge theory can have a deconfinement phase in $3$D where the three generations of neutrinos becomes well defined at low energy.

The same argument also apply to the off-diagonal mass term:
\begin{eqnarray}
\mathcal{L}_{m-od}
&=&\frac{ig_{cd}}{4}\phi(x)\left[\overline {\psi}_d(x)(1+\gamma_0\gamma_5)(1+\gamma_5)\psi_c(x)+\overline {\psi}_c(x)(1+\gamma_5)(1-\gamma_0\gamma_5)\psi_d(x)\right]
\nonumber\\&+&\frac{ig_{cf}}{4}\phi(x)\left[\overline{\psi}_f(x)(1+\gamma_5)\psi_c(x)
+\overline\psi_c(x)(1+\gamma_5)\psi_f(x)\right]\nonumber\\&+&\frac{ig_{df}}{4}\phi(x)\left[\overline{\psi}_f(x)(1-\gamma_0\gamma_5)\psi_d(x)
+\overline\psi_d(x)(1+\gamma_0\gamma_5)\psi_f(x)\right]\label{massod},
\end{eqnarray}
which can be expressed as:
\begin{eqnarray}
 &&\frac{ig_{cd}}{4}\phi(x)\left[\overline {\psi}_d(x)(1+\gamma_0\gamma_5)(1+\gamma_5)\psi_c(x)+\overline {\psi}_c(x)(1+\gamma_5)(1-\gamma_0\gamma_5)\psi_d(x)\right]\nonumber\\&=&
 \frac{ig_{cd}}{2}\phi(x)\left[\gamma_\up(x)\gamma_\up^\prime(x)-\gamma_\down(x)\gamma_\down^\prime(x)\right];
 \end{eqnarray}
 \begin{eqnarray}
 &&\frac{ig_{cf}}{4}\phi(x)\left[\overline{\psi}_f(x)(1+\gamma_5)\psi_c(x)
+\overline\psi_c(x)(1+\gamma_5)\psi_f(x)\right]\nonumber\\&=&
 \frac{ig_{cf}}{2}\phi(x)\left[\gamma_\up(x)\gamma_\up^\prime(x)-\gamma_\down(x)\gamma_\down^\prime(x)\right];
  \end{eqnarray}
 \begin{eqnarray}
 &&\frac{ig_{df}}{4}\phi(x)\left[\overline{\psi}_f(x)(1-\gamma_0\gamma_5)\psi_d(x)
+\overline\psi_d(x)(1+\gamma_0\gamma_5)\psi_f(x)\right]\nonumber\\&=&
 \frac{ig_{df}}{2}\phi(x)\left[\gamma_\up(x)\gamma_\up^\prime(x)-\gamma_\down(x)\gamma_\down^\prime(x)\right],
\end{eqnarray}
Thus we can derive $g_{cd}=g_{cf}=g_{df}=g^\prime$. Finally, by comparing the diagonal and off-diagonal mass terms, we can further derive $|g|=|g^\prime|$. Here the relative sign of $g$ and $g^\prime$ can not be fixed because this relation is not a consequence of $\mathbb{Z}_2\otimes \mathbb{Z}_2$ flavor symmetry(We note that flavor symmetry can \emph{not} relate diagonal and off-diagonal mass terms).

In conclusion, all the above results come from a single principle -- the three generations of neutrinos/anti-neutrinos are the three resonating states out of the \emph{same} four Majorana zero modes at cutoff scale. We conjecture that the $\mathbb{Z}_2\otimes \mathbb{Z}_2$ flavor gauge symmetry proposed here is also crucial for understanding the charged lepton and quark mass hierarchy problem, which might originate from the spontaneously breaking of such a flavor gauge symmetry.

\subsection{Symmetry properties of neutrino mass mixing matrix}
\label{sec: mass symmetry}
 Now let us examine the symmetry of the derived mass mixing matrix. Although the mixing angle derived above is consistent with the GR pattern, the symmetry group is different from that in Ref.\cite{masssymmetry1,masssymmetry2}, and it contains three $\mathbb{Z}_2$ generators $U$, $S$ and $R$, defined by:
\begin{eqnarray}
U=\left(
    \begin{array}{ccc}
      1 & 0 & 0 \\
      0 & 0 & 1 \\
      0 & 1 & 0 \\
    \end{array}
  \right);
S=\frac{1}{\sqrt{5}}\left(
    \begin{array}{ccc}
      1 & -\sqrt{2} & -\sqrt{2} \\
      -\sqrt{2} & -\frac{(\sqrt{5}+1)}{2} & \frac{(\sqrt{5}-1)}{2} \\
      -\sqrt{2} & \frac{(\sqrt{5}-1)}{2}  &-\frac{(\sqrt{5}+1)}{2}\\
    \end{array}
  \right);
  R=\frac{1}{\sqrt{2}}\left(
    \begin{array}{ccc}
      0 & i & i \\
      -i& \frac{1}{\sqrt{2}}  & -\frac{1}{\sqrt{2}} \\
      -i & -\frac{1}{\sqrt{2}} & \frac{1}{\sqrt{2}} \\
    \end{array}
  \right),
\end{eqnarray}
They satisfy:
\begin{eqnarray}
U^TMU=M; \quad S^TMS=M; \quad R^TMR=M,
\end{eqnarray}
and
\begin{eqnarray}
U^2&=&1;\quad S^2=1;\quad R^2=1,\nonumber \\ US&=&SU;\quad UR=RU;\quad SR=-URS,
\end{eqnarray}
$U$ is the center of the above symmetry algebra since it commutes with both $S$ and $R$. The above algebra indeed implies a $D_4$ symmetry.
We note that $U$ and $S$ are the two $\mathbb{Z}_2$ generators of the GR pattern\cite{masssymmetry1,masssymmetry2} characterized by the $\mathbb{Z}_2\otimes \mathbb{Z}_2$ Klein symmetry and apply to generic $g,g^\prime$, while $R$ is a new generator which arises from the special relation $g=-g^\prime$.
\end{widetext}

\bibliography{neutrino} 
\end{document}